\documentclass[conference]{IEEEtran}
\IEEEoverridecommandlockouts
\usepackage{cite}
\usepackage{amsmath,amssymb,amsfonts}
\usepackage{algorithmic}
\usepackage{graphicx}
\usepackage{textcomp}
\usepackage{xcolor}
\usepackage[hyphens]{url}
\usepackage[colorlinks=true, linkcolor=blue, citecolor=blue, urlcolor=blue, breaklinks=true]{hyperref}
\usepackage{booktabs}
\usepackage{tabularx}
\usepackage{threeparttable}

\begin{document}

\bstctlcite{BSTcontrol}

\title{AI Data Centers and Power System Sustainability: Understanding the Sustainability Implications of AI-Driven Data Centers on Power Systems}

\author{\IEEEauthorblockN{Yuhao Huang, Novarun Deb, Hamidreza Zareipour}
\IEEEauthorblockA{\textit{Department of Electrical and Software Engineering} \\
\textit{University of Calgary}\\
Calgary, Alberta, Canada \\
\{yuhao.huang, novarun.deb, hzareipo\}@ucalgary.ca}
}

\maketitle

\begin{abstract}

The rapid expansion of artificial intelligence (AI) has driven unprecedented growth in data center electricity demand. The scale and pace of this load growth carry significant implications for the sustainability of electric power systems. On the one hand, rapid, spatially concentrated data center load growth is outpacing clean energy deployment in several major regions, raising emissions and challenging both grid flexibility and reliability. On the other hand, this fast-developing and capital-intensive sector offers abundant opportunities to advance sustainability through clean energy integration and operational innovations. This article provides an overview of the mechanisms through which data center affect power system sustainability, underscoring both risks and the potential. Specifically, this article (i) characterizes AI data center load behavior and categorizes electricity supply configurations by function and sustainability profile, as well as situates these loads within global and regional electricity demand trends; (ii) analyzes sustainability impacts across short-run operational and long-run planning mechanisms, evaluates effects on grid carbon emissions and renewable energy utilization, and feasibility  of offering system flexibility and participating in ancillary service; and (iii) evaluates real-world corporate sustainability pathways and highlighting both the system benefits and feasibility limits of current carbon accounting practices. The goal of this work is to synthesize existing knowledge and technological developments and to guide research and development toward a more sustainable integration of AI data centers and electric power systems. 

\end{abstract}

\begin{IEEEkeywords}
AI data center, sustainability, power system
\end{IEEEkeywords}

\section{Introduction}
The rapid advancement and widespread deployment of artificial intelligence (AI) have led to an unprecedented surge in global compute demand, driving a new wave of large-scale data center development. As of early 2024, more than 11,000 data centers were registered worldwide, a number that continues to grow as AI applications expand across industries \cite{chalamala_data_2025}. A significant escalation in electricity use accompanies this rapid expansion. This rapid expansion is accompanied by a significant escalation in electricity use. According to International Energy Agency’s (IEA) estimation, global data centers consumed approximately 415 TWh of electricity in 2024, representing around 1.5\% of global demand. Projections indicate that consumption may more than double to 945 TWh by 2030, with AI identified as the principal driver of growth \cite{iea_energy_2025}. In the United States, electricity use by data centers increased from 76 TWh in 2018 to 176 TWh in 2023, representing 4.4\% of national demand, and is expected to reach 325–580 TWh or 74–132 GW by 2028 under different development scenarios \cite{shehabi_2024_2024}.

From a power perspective, the scale of modern AI data centers is striking. Individual hyperscale AI campuses are now being designed with gigawatt-level power requirements, and some planned developments exceed 10 GW \cite{nvidia_openai_2025}, comparable to the peak load of the entire country of Switzerland \cite{iea_switzerland_2023}. Such single-site power demands, rivalling those of major metropolitan regions, illustrate how AI-driven computing has transformed data centers into one of the largest concentrated electrical loads in today’s power systems. At the aggregate level, this trend is equally pronounced: Federal Energy Regulatory Commission (FERC) expects U.S. data center demand to reach nearly 21 GW in 2024, up from 19 GW in 2023, and to grow further to 35 GW by 2030 \cite{ferc_summer_2024}, making AI data centers one of the fastest-growing contributors to national power demand growth. 

The unprecedented scale and rapid growth of AI data center loads have profound implications for the sustainability of electric power systems. As demand rises faster than new clean-energy capacity can be deployed, many regions are experiencing growing pressure to rely more heavily on fossil-based generation to maintain system reliability \cite{chalamala_data_2025,national_academies_of_sciences_engineering_and_medicine_implications_2025}. In Virginia, US, Dominion Energy has indicated in its planning documents that projected data center load growth, in combination with other regional demand drivers and transmission constraints, could require postponing some fossil plant retirements and expanding natural gas generation under certain scenarios, despite the state’s decarbonization mandates \cite{national_academies_of_sciences_engineering_and_medicine_implications_2025}. Similar tensions are also emerging elsewhere, as IEA projections show that more than half of the additional electricity required for data centers through 2035 will come from non-renewable sources, with natural-gas generation alone increasing by 220–285 TWh to meet rising demand in key markets such as the United States and the Middle East \cite{iea_world_2025}. As a result, the rapid expansion of AI data centers risks outpacing the clean-energy transition, potentially increasing system-level carbon emissions, constraining renewable integration, and creating new sustainability challenges for power grids worldwide. 

While significant challenges exist, the abundant opportunities brought by this technology surge are undeniable. The operators of advanced and capital-intensive AI data centers, often global leading technology companies, tend to adopt emerging technologies earlier than many other sectors and have shown a willingness to pursue innovative pathways for reducing their environmental footprints \cite{meta_2025_2025, google_environment_2025, amazon_2024_2024, microsoft_2025_2025}. The nature of AI workloads provides operational flexibility and grid integration diversity that is uncommon among traditional large industrial loads, thereby enhancing renewable penetration, reducing renewable curtailment, and smoothing net load variability \cite{chen_electricity_2025}. For example, some data center projects invest in advanced energy-management systems \cite{ben_gomes_our_2025, agarwal_redesigning_2021}, adopt large-scale green energy storage systems \cite{meta_2025_2025, matthew_gooding_microsoft_2024}, and procure clean energy through diverse contracting mechanisms \cite{meta_2025_2025, amazon_2024_2024, google_powering_nodate}.

Motivated by this duality, this paper examines the integration pathways and environmental sustainability impacts of AI data centers on power systems. The analysis begins by characterizing AI data center load behavior and situating these loads within global and regional electricity-demand trends, alongside other major electrification drivers. It then establishes baselines for evaluating sustainability impacts. Building on this foundation, the article analyzes AI data center’s short-run operational and long-run structural effects on grid carbon emissions, renewable energy utilization, and system flexibility, while accounting for regional variation and key sources of uncertainty. The discussion then examines the potential system benefits associated with workload flexibility, participation in ancillary service and demand response markets, and the role of data centers as anchor loads for renewable and transmission investment. Finally, the article reviews corporate clean energy procurement practices and evaluates both the contributions and the limitations of prevailing carbon accounting frameworks.  

This article focuses on the use-phase environmental impacts of AI data centers from the perspective of the electric grid, examining how incremental load affects grid carbon emissions, renewable energy utilization, and system flexibility. The analysis explicitly excludes internal data center efficiency technologies and life-cycle emissions associated with equipment manufacturing and end-of-life processes, as well as broader dimensions of sustainability such as economic equity, affordability impacts, and social or community outcomes. These topics are essential, however fall outside the analytical scope of this work, which aims to provide a rigorous and system-level assessment of environmental mechanisms related to the power sector. 

\section{AI Data Center Loads: Characteristics and Context}

\subsection{Load characteristics of AI data centers}

AI-driven data centers differ from traditional IT facilities in that they are purpose-built to support AI-related workloads. These facilities deploy clusters of AI-dedicated graphics processing units (GPUs), tensor processing units (TPUs), and high-performance central processing units (CPUs) that enable large-scale parallel computing, and this architecture leads to much higher power density than conventional data centers \cite{shehabi_2024_2024, north_american_electric_reliability_corporation_characteristics_2025}. Whereas traditional IT facilities generally consumes 7–10 kW per rack, AI data centers can reach a power density level of 30–100 kW \cite{epri_powering_2024}. In addition to higher power rating, AI data centers demonstrate more volatile load behavior. The bursty nature of AI workloads can cause operational power to fluctuate by megawatt level within a second, a behavior rarely seen in traditional facilities \cite{chalamala_data_2025, chen_electricity_2025, north_american_electric_reliability_corporation_characteristics_2025}.  

The pattern of load variation also depends on the stage of the AI pipeline. Training is generally the most electricity-intensive phase, during which power demand remains consistently high due to intensive forward and backward computations. Sharp fluctuations in load can occur during checkpointing, when computation temporarily pauses and parameters are written to storage, followed by a rapid return to high usage once training resumes \cite{chalamala_data_2025, chen_electricity_2025, north_american_electric_reliability_corporation_characteristics_2025}.  

\begin{figure*}[t]
    \centering
    \includegraphics[width=\linewidth]{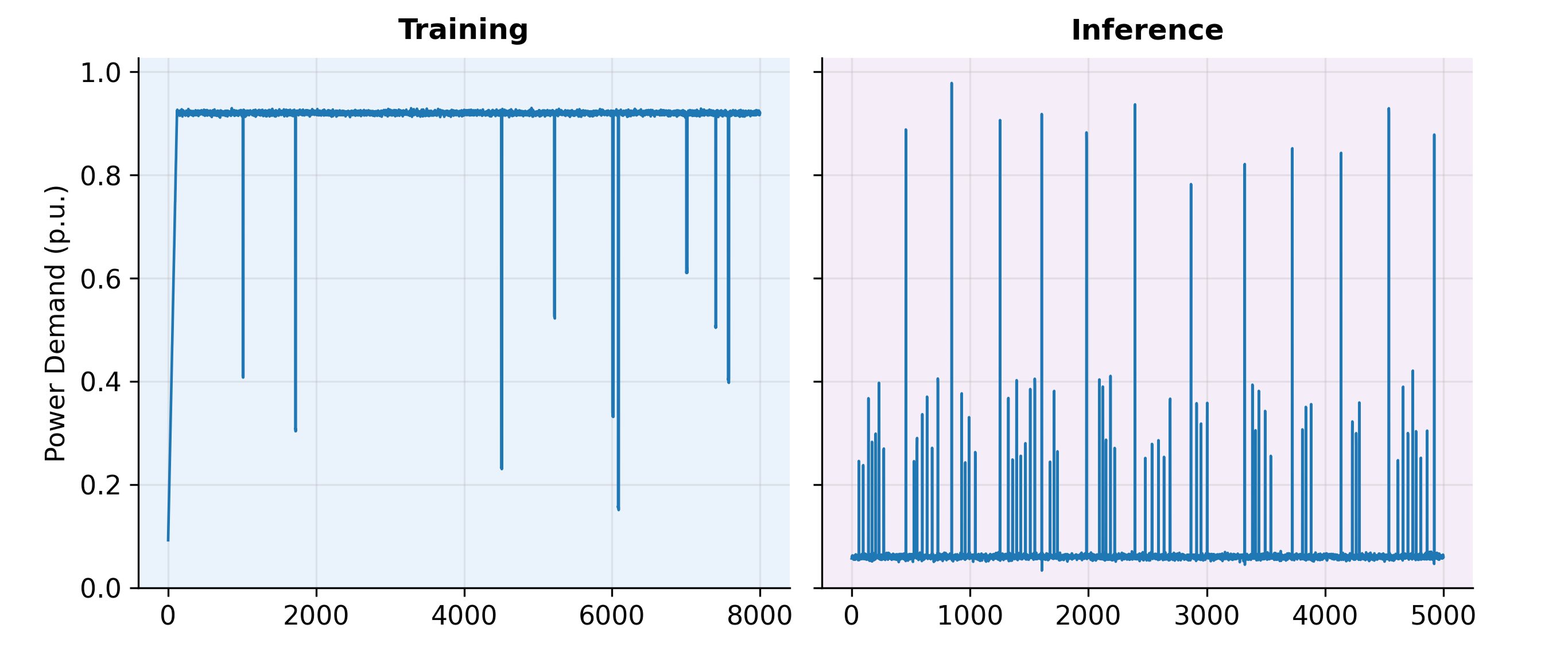} 
    \caption{Conceptual illustration of power demand patterns of AI load during training and inference stages. The demand profiles shown in Figure 1 are simulated based on patterns reported in \cite{liUnseenAIDisruptions2024} and are intended to highlight qualitative load characteristics rather than represent measured power traces. The horizontal axis represents relative time scale rather than measured seconds, and power is shown in per-unit (p.u.) values. }
    \label{Fig1}
\end{figure*}

Inference, although less power-intensive per query, can account for up to 90\% of an AI model’s total lifecycle energy use when serving billions of user requests \cite{touvron_llama_2023}. Inference loads often exhibit relatively low average power, along with bursty spikes due to user activity and unpredictable task complexity, and clear daily usage cycles with peak demand aligned with business and social activity hours \cite{chen_electricity_2025}. It is also common for AI data centers to be purpose-built for either training or inference \cite{north_american_electric_reliability_corporation_characteristics_2025}. Taken together, these properties make AI data centers not only larger electricity consumers but also inherently more dynamic and operationally complex loads for the power system. Figure \ref{Fig1} presents an illustrative comparison of power demand profiles for AI training and inference workloads. To emphasize qualitative differences in load patterns, the horizontal axis represents relative time scale rather than measured seconds, and power is shown in per-unit (p.u.) values.

\subsection{Supply Mix of AI data centers}

\begin{figure*}[t]
    \centering
    \includegraphics[width=\linewidth]{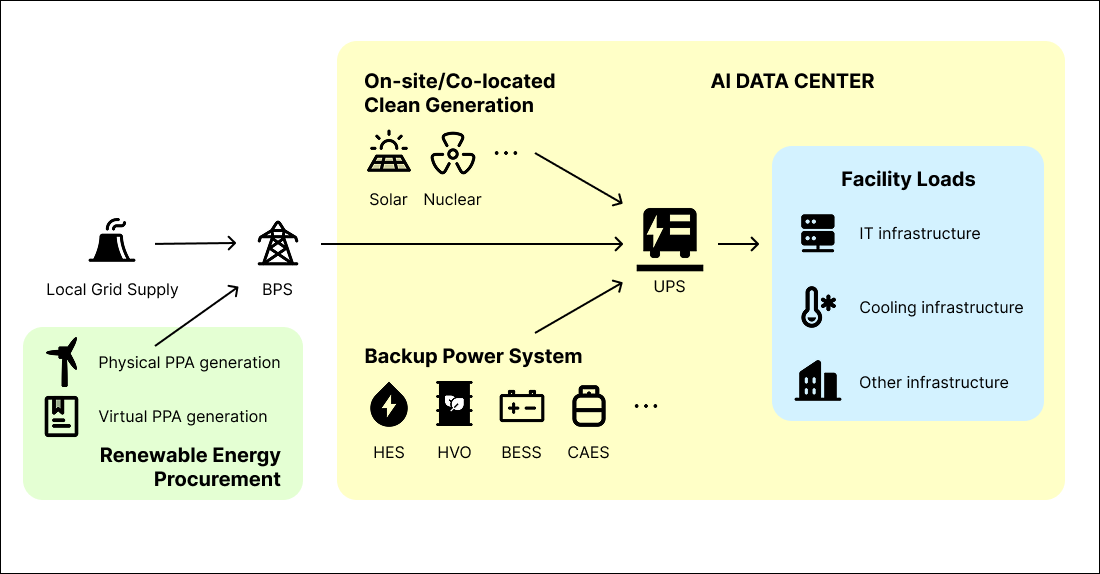} 
    \caption{System level architecture of the electricity mix for AI data centers, including grid supply, onsite and co-located clean generation, renewable energy procurement, and backup power systems. AI data centers draw electricity from multiple categories of supply resources. Grid supplied electricity is delivered through the Bulk Power System (BPS), including transmission and distribution networks. Onsite or co-located clean generation includes technologies such as solar photovoltaic systems and nuclear generation, among other low-emitting sources. Renewable energy procurement refers to contractual mechanisms such as Physical and Virtual Power Purchase Agreements (PPA) used to support the development or acquisition of clean electricity. Backup systems provide resilience during outages and include generator-based solutions such as Hydrotreated Vegetable Oil (HVO), and energy storage technologies such as Hydrogen Energy Storage (HES), Battery Energy Storage Systems (BESS) and Compressed Air Energy Storage (CAES). }
    \label{Fig2}
\end{figure*}

AI data centers rely on a combination of electricity sources to support their uninterrupted and highly power-intensive operation \cite{iea_energy_2025}. Figure \ref{Fig2} illustrates a system-level architecture of AI data center power supply mix. The electricity supply mix can be grouped into three main categories: grid-supplied electricity, procured or onsite clean energy resources, and backup and resilience-oriented systems \cite{chen_electricity_2025, yu_global_2024}. Here, we refer to clean energy as energy sources that produce little to zero climate-warming greenhouse gas (GHG) emissions during operations, including but not limited to solar, wind, hydropower, bioenergy, and nuclear. Overall, these three categories differ not only in technical function but also in the ways they shape the sustainability profile of both data centers themselves and power system, for example, through mechanisms including carbon intensity, renewable-energy integration, and system flexibility. Notably, in this article, system flexibility refers to the ability of the energy system to adjust generation and consumption in response to signals, including market change, grid conditions, and other external factors, to maintain reliable and cost-efficient operation of the grid.

\subsubsection{Grid Power Supply }

Most AI data centers are connected to the bulk power system (BPS), either through medium- or high-voltage distribution networks or directly at the transmission level, depending on the scale of their power demand \cite{chen_electricity_2025, north_american_electric_reliability_corporation_characteristics_2025, yu_global_2024}. As a result, the grid carbon intensity determined by the generation mix and power flow is the primary determinant of a data center’s operational cleanliness \cite{maji_data_2025}. The interaction, however, is bidirectional. As many power systems increase renewable penetration and reduce carbon intensity in pursuit of climate goals \cite{ember_yearly_2025}, AI data centers indirectly benefit from cleaner electricity, leading to reductions in their average operational emissions. Conversely, at the same time, AI data centers often cluster geographically, and rapid load growth in these regions may outpace local clean energy capacity addition or transmission expansion \cite{national_academies_of_sciences_engineering_and_medicine_implications_2025, iea_world_2025, epri_powering_2024}. Under such circumstances, system operators may rely more frequently on existing marginal units, which are often fossil-based \cite{iea_energy_2025, national_academies_of_sciences_engineering_and_medicine_implications_2025}. The dynamics behind this are examined in greater detail in the subsequent ‘Long-Run Structural Impact’ section and are thus not elaborated here. 

\subsubsection{Onsite and Procured Clean-Energy Resources}

In addition to drawing electricity from the grid, AI data centers are increasingly applying a variety of clean-energy solutions to enhance their sustainability. One prominent strategy is co-location with dedicated low-carbon energy resources \cite{chalamala_data_2025, chen_electricity_2025}, such as solar photovoltaics \cite{morohubWorldsLargestGreen2025} and nuclear plants \cite{michael_terrell_new_2024}. These sources can provide stable low-carbon power directly to high-demand AI workloads while reducing dependence on grid transmission and easing local grid-interconnection constraints \cite{chen_electricity_2025}. A notable example is the 2024 strategic partnership between Intersect Power, Google and TPG Rise Climate, which aims to co-locate gigawatt-scale data center load with newly built renewable power generation and battery storage facilities \cite{intersect_intersect_2024, ruth_porat_new_2024}. 

Beyond directly financing physical generation assets, another major way for AI data centers to procure clean electricity is through power purchase agreements (PPAs) \cite{kansal_introduction_2018}, which support the development of renewable projects by dedicated energy developers. Depending on the contract structure, PPAs may take the form of either physical or virtual agreements. In physical agreements, clean energy generated locally is delivered to data centers through the BPS transmission and distribution network. In contrast, under virtual agreements, clean energy is generated at a remote location and used primarily for contractual settlement and firm-level carbon accounting rather than direct physical supply \cite{kansal_introduction_2018}. Such virtual procurement mechanisms have become the dominant solution among major technology firms, many of which report gigawatt-scale renewable contracting in their annual sustainability disclosures \cite{meta_2025_2025, google_environment_2025, amazon_2024_2024, microsoft_2025_2025}, while physical PPAs remain comparatively less common \cite{kansal_introduction_2018}. However, the actual environmental impact of these strategies depends on the temporal and locational alignment between clean energy output and real-time data center demand \cite{iea_energy_2025, national_academies_of_sciences_engineering_and_medicine_implications_2025}. Mismatches in timing or geography can limit the extent to which procured clean energy displaces local fossil generation, an issue examined further in the section ‘From Annual Matching to Granular Carbon Accounting: Practices, Limitations, and Challenges.’ 

\subsubsection{Backup and Resilience Systems}

To ensure uninterrupted operation, AI data centers typically deploy backup power systems, including uninterruptible power supply (UPS), and generator- or battery-based energy resources. The UPS serves as a central power-conditioning and routing interface during normal operation. While not a primary power source, it provides backup power and energy buffering during short-duration primary supply interruptions and is therefore commonly categorized as part of data center power supply system. The instantaneous power support maintains continuous operation during a primary supply outage, providing sufficient time for the main source to recover or for backup generators to start and re-energize the load \cite{aamir_review_2016}. During extended outages, other backup power sources come into play. Historically, diesel backup generators have remained the industry baseline due to their technological maturity, high energy density, and fast start-up capability \cite{kambhampati_moving_2024}. However, concerns over carbon emissions and local air pollution issues have motivated a shift toward greener alternatives, including hydrotreated vegetable oil (HVO), hydrogen fuel cells, advanced BESS, and compressed air energy storage (CAES) \cite{kambhampati_moving_2024}. Each technology offers different backup durations and environmental footprints.  

Hydrogen fuel cells have demonstrated their feasibility at hyperscale: in 2023, Microsoft successfully powered a data center for 48 hours using a 1.5-MW hydrogen fuel cell system, marking a notably extended backup operation \cite{matthew_gooding_microsoft_2024}. HVO is also being adopted as a drop-in low-carbon diesel replacement; Meta’s Clonee data center in Ireland was the first to transition fully, and the company expanded HVO as a backup source to additional North American facilities in 2025 as part of its strategy to reduce backup-generation footprint \cite{meta_2025_2025}. BESS are likewise emerging as a lower-emission option for numerous operators. Meta, for instance, is deploying a 50-MW, four-hour BESS alongside the Sky Ranch Solar Energy Center in New Mexico to provide stored clean energy that can support data center operations during grid disruptions \cite{pnm_pnm_2024}. For longer-duration needs, CAES has also been explored. Large installations such as the Huntorf and McIntosh plants—both capable of multi-hour discharge with high operational reliability—illustrate the potential suitability of CAES for data center resilience in regions with appropriate geological formations \cite{he_exergy_2017}. 

\subsection{Macroscale Growth of AI Data Center Loads}

Understanding the power system implications of AI data centers requires situating their growth within the broader trajectory of global data center electricity demand. This section synthesizes the latest quantitative projections from international agencies, industry analyses, and national regulators to provide a macroscale view of data center load growth. Throughout, the projections refer to all data center electricity consumption, since AI-specific data centers are not separately measured in most statistical systems. As a practical proxy, however, IEA estimates that the electricity consumption of accelerated servers accounted for roughly 24\% of global server electricity use and about 15\% of total data center electricity demand in 2024 \cite{iea_energy_2025}. These shares are expected to rise substantially as AI training intensity scales and model sizes increase, underscoring the uncertainty about future AI-specific data center electricity demand. 

Recent projections indicate that total data center electricity consumption is entering a period of rapid acceleration. In the Energy and AI report, IEA has established future projections under a base case scenario, which takes the current projection for server shipments to 2028 into further trajectory \cite{iea_energy_2025}. Under the base case scenario, global data center electricity consumption rises from 415 TWh in 2024 to 945 TWh in 2030, corresponding to a 14.7\% compound annual growth rate. By contrast, electricity uses from accelerated servers grows at approximately 30\% per year, far outpacing both the total data center load and the 9\% annual growth of conventional server demand. This asymmetry illustrates the disproportionate contribution of AI workloads to future data center electricity growth.  

The globally aggregated trend is however not universal in every region or country. In the IEA base case scenario, up to 2024, more than 85\% of global data center electricity consumption is located in three regions: the US ($\approx45\%$), China ($\approx25\%$), and Europe ($\approx15\%$), and they together are expected to account for nearly 87\% of the global increase through 2030 \cite{iea_energy_2025}. In the US, data center electricity consumption is expected to increase by roughly 240 TWh between 2024 and 2030, representing a 130\% expansion and making it one of the fastest-growing load categories in the country. China is projected to add approximately 175 TWh in data center demand over the same period, a 170\% increase, ranking alongside air conditioning and electric vehicle charging as one of the three major contributors to national demand growth. Europe, despite a smaller absolute scale and slightly slowing down growth rate, is still expected to see data center electricity use increase by 45 TWh, or 70\%, by 2030.  

While projections emphasize demand growth, the sustainability implications depend critically on regional power system conditions. The three regions above that dominate global data center electricity use differ substantially in generation mix, transmission capacity, planning institutions, and marginal emission profiles. 

\begin{figure*}[t]
    \centering
    \includegraphics[width=\linewidth]{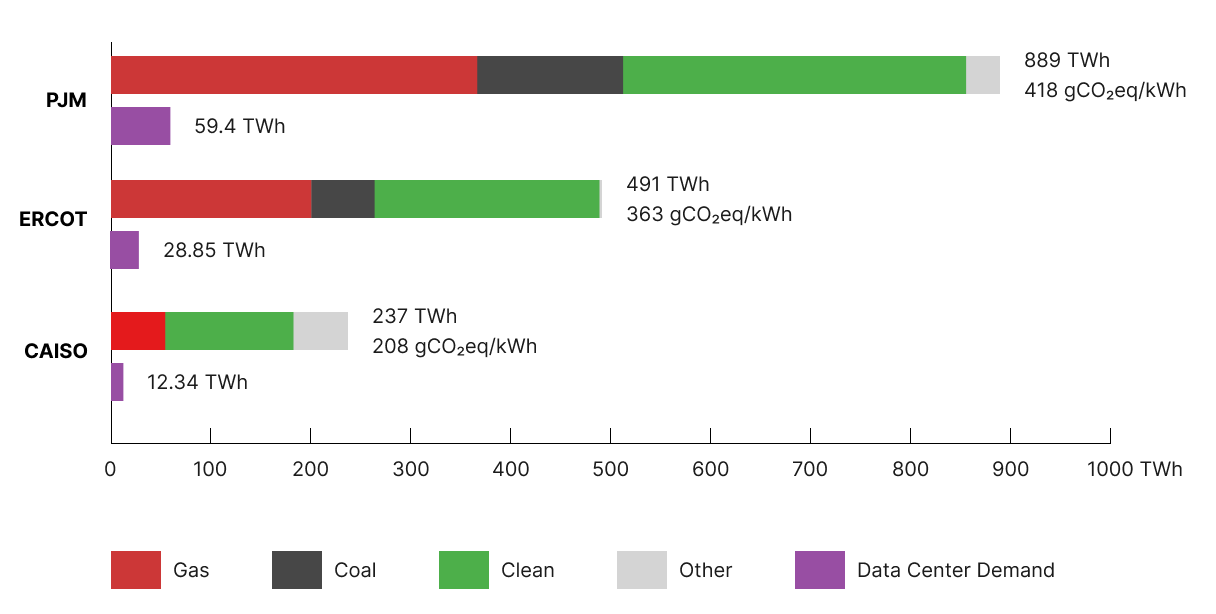} 
    \caption{Electricity generation mix and data center demand across three major US RTOs in 2025. Generation mix and carbon intensity values for PJM, ERCOT, and CAISO correspond to 2025 electricity generation and are sourced from Electricity Maps \cite{zotero-item-797, zotero-item-796, zotero-item-798}. Clean energy includes renewable technologies such as hydro, solar, wind, and geothermal as well as biomass and nuclear. Data center electricity demand projections are taken from the IM3+EPRI Data Center Load Projections dataset \cite{burleysonIM3EPRIData}, which projects annual state level data center demand based on observed 2023 loads under multiple growth scenarios. This figure uses the 2025 projection under the 15\% annual growth scenario, which aligns most closely with growth rates implied by IEA forecasts. For PJM, data center demand is obtained by aggregating projected demand for states with a non-zero data center demand and for which PJM covers the majority of load service territory, including Maryland, New Jersey, Ohio, Pennsylvania, and Virginia \cite{zotero-item-801}. }
    \label{Fig3}
\end{figure*}

\textbf{United States}. The US power system is characterized by regional heterogeneity. In two of the major Regional Transmission Organizations (RTO) in the country, Pennsylvania-New Jersey-Maryland Interconnection (PJM) and Electric Reliability Council of Texas (ERCOT), natural gas and coal represent a large share of total generation, with low-carbon resources penetration continuously rising \cite{USElectricity2025SpecialReport}. In contrast, California ISO (CAISO) operates a system with a lower fossil-fuel share and correspondingly higher penetration of low-carbon resources \cite{zotero-item-768}. Figure 3 presents a comparison of the generation mix and data center electricity demand in the three RTOs for 2025. Furthermore, significant interconnection congestion is observed across several regions. Notably, Virginia and Texas, the two regions expecting most fastest-growing data center demand in the US \cite{epri_powering_2024}, lie within PJM and ERCOT, respectively, followed by California, which is covered by CAISO. Rapid AI-driven load growth is emerging precisely in markets where marginal generation is predominantly gas-fired, and transmission constraints already shape operational and investment outcomes. 

\textbf{China}. China’s electricity supply is dominated by coal generation, with renewable penetration such as solar and wind rising rapidly \cite{20260210_CN2025monthly}. Structural features such as long-distance west-to-east transmission \cite{sunDevelopmentUHVPower2017}, regional imbalance between generation and load centers, and curtailment in resource-rich northern provinces shape the marginal emissions associated with incremental load \cite{weiRegionalDisparitiesVariation2025}. These factors imply that AI data center growth in coastal load centers may increase coal-based marginal supply unless local renewable build-out accelerates. 

\textbf{Europe}. Europe operates with high renewable penetration, strong cross-border interconnections, and increasingly stringent decarbonization mandates \cite{europeanenvironmentagency.TrendsProjectionsEurope2025}. Yet marginal emissions vary sharply across countries: Nordic systems offer low-carbon marginal supply \cite{Nordicenergyresearch202405}, while gas-fired generation dominates in parts of Western and Southern Europe \cite{zotero-item-784}. Transmission congestion and rising peak-demand constraints \cite{neumannPotentialRoleHydrogen2023} have tightened capacity margins in the UK, Ireland, and the Netherlands, where data center concentration is high. 

Beyond advanced economies, emerging data center markets in developing regions are also shaping the global AI-driven electricity demand, each within distinct power system contexts. The IEA base case scenario suggests that, in India, rapid digitalisation and data-localisation policies have driven a doubling of total installed data center capacity over the past few years and put the sector on track to approach 5 GW by 2030 \cite{iea_energy_2025}, even as coal continues to supply the majority of grid electricity and reliability challenges lead operators to rely on backup generation and grid upgrades to serve new load. As a result, incremental data center electricity demand in India is more likely to depend on existing fossil-dominated generation while renewable and transmission investment catches up. Elsewhere, the Middle East is also experiencing a fast-growing data center investment cycle supported by sovereign digital-economy strategies and hyperscale AI infrastructure initiatives \cite{zotero-item-792, zotero-item-793}. The IEA base case scenario projection estimates the total data center electricity consumption in the region to reach 3 TWh by 2030, doubling the consumption in 2024 \cite{iea_energy_2025}. Major cities in the region such as Dubai are positioning themselves as regional digital hubs that link Europe, Asia, and Africa. In such environments, abundant energy resources and economic diversification agendas interact with nascent grid planning frameworks, producing power system conditions that differ markedly from those in advanced markets.

\subsection{AI Data Center Loads in the Context of Broader Electrification}

The rapid expansion of AI data centers is unfolding within the broader global shift toward electrification \cite{iea_world_2025}, making it important to evaluate their growth relative to other major emerging electricity loads such as electric vehicles (EVs) and heat pumps.  

 In absolute terms, data centers are not the only dominant source of new electricity demand. IEA projects that global data center electricity consumption will rise from 416 TWh in 2024 to 946 TWh in 2030 \cite{iea_energy_2025}, while electricity demand from EVs will increase from 180 TWh to 780 TWh over the same period \cite{GlobalEVOutlook2025a}. By comparison, the growth associated with heat pumps is smaller. IEA projects that heat pump related electricity demand in the US, China, and the European Union (EU) will increase by a combined 125 TWh between 2025 and 2030 \cite{Electricity_2026}. In overall terms, the base case scenario suggests that data center load growth accounts for less than 10\% of total global electricity demand growth \cite{iea_energy_2025}. 

Even so, the power system significance of data centers is distinctive and cannot be assessed by TWh growth alone. Unlike EVs and heat pump demand, which are relatively dispersed across end users and often more weather or behavior dependent, data center demand is highly concentrated geographically, frequently appears as very large single site loads, and typically operates with high load factors and strong temporal persistence. In addition, data centers are dominated by power electronic equipment and can exhibit rapid ramping, non-linear load behavior, and demanding requirements for power quality and reliability. Their internal backup systems, sensitivity to voltage disturbances, and limited tolerance for interruption further distinguish them from more flexible loads such as hydrogen electrolysis. As a result, even though data centers contribute a modest share of aggregate electricity demand growth, their concentrated, continuous, and technically complex load profile can create uniquely strong effects on transmission expansion, local resource adequacy, balancing requirements, and grid stability. In this sense, the grid impact of AI driven data center growth is structurally different from that of broader electrification loads. 
\section{Impacts on Grid Sustainability}

\subsection{Baselines for Assessing Sustainability Impacts}

Assessing the sustainability implications of AI data center growth requires a clear definition of the baselines against which both operational marginal effects and long-term structural impacts can be interpreted. First, historical and current average carbon intensity provides a foundational systemwide reference point. Based on IEA Electricity 2025 report, global power sector $\mathrm{tCO_2}$ emissions in 2024 were approximately 13.8 Gt, with electricity generation approaching 30,000 TWh \cite{iea_electricity_2025}, corresponding to a global average carbon intensity of roughly 460–480 $\mathrm{gCO_2eq/kWh}$. While this metric reflects the overall cleanliness of power sector, it does not capture how emissions change at the margin when additional load is added. 

For operational analysis, marginal rather than average emission intensity is the relevant indicator. Empirical estimates indicate that gas-dominated systems typically exhibit marginal $\mathrm{tCO_2}$ intensities on the order of 400–600 $\mathrm{gCO_2eq/kWh}$, while coal-dominated systems frequently exceed 700–900 $\mathrm{gCO_2eq/kWh}$, depending on technology, heat rates, and fuel conditions \cite{hawkesEstimatingMarginalCO22010, siler-evansMarginalEmissionsFactors}. In systems with high renewable penetration, short-run marginal emissions can be very low during periods of renewable surplus, because supplying incremental demand does not require increasing fossil output. For example, in CAISO during mid-day periods when solar provides around 60\% of available electricity, the average grid carbon intensity can fall to roughly 60 $\mathrm{gCO_2eq/kWh}$ based on recent operational observations from early March, 2026 \cite{zotero-item-802}. At these times solar generation itself has an operational emission intensity of about 26 $\mathrm{gCO_2eq/kWh}$ and natural gas contributes only around 3\% of electricity supply \cite{zotero-item-802}. Since real time marginal emission factors are difficult to observe directly in operational grid data, this example instead illustrates system conditions under which the marginal emissions associated with additional demand are likely to be very low.  

To evaluate long-term structural effects, we adopt the IEA Electricity 2025 projections for the electricity supply needed to meet data center load growth as the global level quantitative baseline. In the Stated Polices Scenario (STEPS) projection, global data center demand through 2035 is met 45\% by renewables, 25\% by natural gas, and 20\% by nuclear energy \cite{iea_electricity_2025}. These trajectories offer a direct quantitative framework to interpret the regional analysis in the following sections. 

\subsection{Grid Emission Impact Mechanisms }

Building on the baselines, the impacts of AI data centers on grid sustainability can be understood through two primary pathways: short-run operational dynamics determined by marginal dispatch conditions, and long-run structural changes shaped by resource planning, infrastructure development, and system investment decisions. 

\subsubsection{Short-Run Operational Impacts }

In the short term, we do not consider addition of new generation resources. In this case, incremental electricity demand from AI data centers is met by the marginal generating unit dispatched to balance the grid at each hour. Consequently, the emissions associated with a given megawatt-hour of data center consumption are determined by real-time system conditions rather than by annual average grid intensity. Within a single region, marginal emissions can vary by factors of two to four across seasons or even between adjacent hours, depending on residual load, renewable availability, transmission constraints, and congestion patterns. 

A simple illustrative calculation highlights the scale of these differences. An incremental 1 MW of AI facility load operating at a 90\% utilization factor consumes approximately 21.6 MWh of electricity daily. If served predominantly by coal-based marginal generation at 800 $\mathrm{gCO_2eq/kWh}$, this load would be associated with roughly 17.3 $\mathrm{tCO_2}$ per day. Under gas-based marginal conditions at 500 $\mathrm{gCO_2eq/kWh}$, daily emissions fall to about 10.8 $\mathrm{tCO_2}$, while in a high-renewable system with marginal intensities near 100 $\mathrm{gCO_2eq/kWh}$, the figure further declines to approximately 2.16 $\mathrm{tCO_2}$. These outcomes, spanning more than 15 $\mathrm{tCO_2}$ on a daily basis, demonstrate that the emissions impact of a given AI workload is largely determined by the carbon characteristics of the marginal generation serving it. The presence of the load itself does not predetermine high emissions, but the surrounding system context does.  

Since in most electricity markets, the marginal supply is typically provided by natural gas or coal-fired generators, incremental load can increase the frequency with which these fossil units set the marginal price, and thus the local emission factor is likely to increase due to the addition of data center load.

\subsubsection{Long-Run Structural Impacts}

Over longer horizons, sustained growth in data center load can influence how grid operators plan for resource adequacy, transmission expansion, and the timing of generator retirements \cite{north_american_electric_reliability_corporation_characteristics_2025, epri_powering_2024}. Because generation, transmission, and interconnection assets have multi-decade lifetimes, sustained AI-driven load growth reshapes the future electricity system in ways that directly influence regional carbon trajectories. 

In many jurisdictions, hyperscale data centers have become one of the dominant contributors to long-term electricity-demand growth, alongside the electrification of transport, heating, and industry. IEA’s Electricity 2025 STEPS scenario reflects this shift, showing substantially higher long-term demand forecasts relative to previous editions and explicitly attributing a significant portion of this acceleration to data centers \cite{iea_electricity_2025}. Importantly, the IEA finds that incremental electricity supply serving data center demand will come from a mix of low-carbon sources and natural gas: in the STEPS trajectory, renewables provide roughly 45\% of incremental supply for data center load through 2035, while gas-fired generation still increases by 220–285 TWh to meet rising data center demand \cite{iea_electricity_2025}. This outlook illustrates that even under expected clean energy expansion, fossil generation remains a substantial component of the marginal supply serving global AI data center growth. 

At the regional level, these imbalances can manifest differently depending on planning horizons, existing generation portfolios, and local transmission constraints. In areas where rapid load growth coincides with limited clean energy build-out or congested interfaces, system planners may anticipate a greater reliance on existing dispatchable resources during certain periods. In jurisdictions where coal or gas retirements had previously been contemplated, accelerated load growth including data centers can prompt reassessments of retirement timelines or consideration of new firm-capacity resources within established planning processes \cite{national_academies_of_sciences_engineering_and_medicine_implications_2025}. Recent developments in several US markets illustrate how these dynamics appear in planning discussions. For example, Entergy Louisiana has proposed more than 2 GW of new natural-gas capacity in connection with serving Meta’s planned 2-GW campus \cite{entergy_entergy_2025}, presenting this as part of a broader reliability and resource adequacy strategy that could span multi-decades. Examples like this reflect how utilities and system operators frame large, geographically concentrated loads within their long-term resource planning narratives, and illustrate how data center growth interacts with preexisting infrastructural, regulatory, and planning conditions that jointly shape long-run capacity decisions. 

\subsubsection{Uncertainty and Regional Variation}

Significant uncertainty surrounds these mechanisms, and any assessment of future impacts must acknowledge their contingent nature. Renewable build-out rates and interconnection timelines remain among the largest sources of uncertainty. Market design features such as capacity accreditation rules, scarcity pricing, and flexibility markets further mediate how incremental demand translates into operational and investment responses. Even where renewable capacity additions exceed total load growth, local congestion or storage undersupply may elevate marginal emissions for extended periods. On longer horizons, planning and regulatory delays may cause fossil units to remain online longer than projected in the IEA baseline, or conversely, targeted clean energy procurement may drive decarbonization faster than expected. Geographic concentration of data center development further amplifies locational effects, as marginal emissions and planning constraints vary sharply across regions. Finally, uncertainty in the evolvement of AI data centers workload, such as the compute intensity of model training, utilization cycles, and cooling requirements, also creates additional variability in both magnitude and timing of incremental demand. 

For these reasons, emission impacts should be interpreted not as deterministic outcomes but as scenario-dependent trajectories determined jointly by system conditions, planning institutions, and the operational characteristics of AI workloads. This framework provides a rigorous foundation for the subsequent analysis of renewable utilization and system flexibility.

\subsection{Flexibility-Driven Impacts on Renewable Utilization}

AI data centers differ from many traditional industrial and commercial loads in that a meaningful portion of their computational demand is inherently flexible \cite{chen_electricity_2025, north_american_electric_reliability_corporation_characteristics_2025}. In practice, flexibility is highly workload dependent. User-facing and latency-critical services such as online inference and interactive applications are largely inflexible because response-time targets and service-level objectives (SLOs) constrain when and where computation can occur. By contrast, batch-oriented workloads associated with the model lifecycle, such as data preparation, and model training and fine-tuning, often tolerate delays within a bounded completion window, creating opportunities to shift demand toward hours and locations with higher renewable availability or lower marginal emissions.  

The values of such flexibility have been documented in both academic and industry studies. A key set of quantitative insights comes from the spatio-temporal load-shifting model developed in \cite{riepin_spatio-temporal_2025}. They examine how assumed flexibility levels (defined as the percentage of allowable deviation between requested and dispatched load in each hour) affect the cost of achieving 24/7 carbon-free matching. Their analysis reveals a consistent trend across 56 data center location combinations in eight European countries: each percentage point of flexible load reduces 24/7 Carbon Free Energy (CFE) costs by 1.29 ± 0.07 €/MWh, with diminishing marginal benefits at higher flexibility levels. These results indicate that even modest flexibility can materially lower the resource requirements of granular CFE matching and that the value of flexibility is robust across diverse siting conditions. 

Evidence from production-scale hyperscale operations reinforces the practical significance of such flexibility. A notable example is Meta’s MAST scheduling framework \cite{choudhury_mast_2024}. Although not a carbon-aware system itself, MAST demonstrates that hyperscale training workloads are already highly mobile across regions. By reducing the GPU demand-to-supply ratio of the most congested region from 2.63 to 0.98, MAST shows that a significant portion of training jobs can be shifted geographically. This evidence supports the practical feasibility of routing flexible AI workloads toward regions and hours with lower marginal emissions, which forms the prerequisite for spatial carbon-aware optimization. 

In the context of power system sustainability, the significance of such flexibility lies in its ability to reshape the marginal emissions profile of data center electricity consumption. The shifting capabilities allow flexible workloads to migrate from hours dominated by fossil-fuel marginal generators to hours with abundant wind or solar generation, and from carbon-intensive regions to grids with higher renewable penetration, thereby producing substantial reductions in marginal emissions without increasing total demand. Flexibility also mitigates renewable curtailment by absorbing excess wind and solar output that would otherwise be unused, improving renewable utilization and reducing system-wide reliance on peaking units. Together, these mechanisms illustrate why workload flexibility is not merely an operational characteristic but a meaningful lever for reducing the carbon footprint of AI compute growth. 

Despite the documented capabilities, operational constraints limit the realizable flexibility range. Radovanović et al. from Google note in \cite{radovanovic_carbon-aware_2023} that flexible workload characteristics, including arrival patterns, resource requirements, and inter-job dependencies, are uncertain and difficult to predict, complicating carbon-aware scheduling at scale. Scheduling decisions must also respect reliability principles, machine capacity limits, and infrastructure protections such as circuit breakers. In spatial scheduling, \cite{choudhury_mast_2024} identifies data–GPU co-location constraints and limited cross-region bandwidth as key barriers to relocating training jobs freely across regions in the MAST scheduler. These factors indicate that while flexibility exists and has been operationalized, it remains conditional on workload structure, infrastructure limits, and system-level coordination requirements. 

\subsection{Feasibility of Ancillary Service and Demand Response Participation}

AI data centers are increasingly equipped with on-site energy storage, primarily through uninterruptible power supply (UPS) systems and large-scale battery energy storage systems (BESS), among the numerous energy solutions introduced in the section “Backup and Resilience Systems”. These resources, originally deployed for reliability, can also contribute to power-system sustainability by providing ancillary services that directly support renewable-energy integration. By absorbing surplus solar or wind generation during periods of oversupply and discharging during peak-demand hours, these systems can both reduce renewable curtailment on the grid and displace high-carbon peaking generators, flattening system load profiles in the process \cite{castillo_battery_2023}. In addition, BESS can supply fast ramping and short-term reserves that help the grid manage variability under high renewable penetration. However, despite these conceptual capabilities and benefits, participation of hyperscale data centers in ancillary service markets is further shaped by a number of practical constraints. Many system operators require precise baselines, verified response performance, sub-minute reaction times, and specific metering configurations that large compute facilities may not readily satisfy under current operational designs. In addition, regulatory frameworks in several US and international markets either limit load participation in certain frequency regulation products or impose minimum participation thresholds that can be difficult for data centers to meet consistently given their high-availability obligations, as well as the significant economic opportunity costs that arise when deferring, interrupting, or relocating compute tasks. These considerations suggest that ancillary service participation is feasible in principle but highly dependent on evolving market structures and operational arrangements.  

AI data centers can also participate in demand response (DR) programs by leveraging the temporal flexibility of AI workloads. When AI data centers strategically reduce or shift load during periods of system stress, they help reduce the dispatch of high carbon peaking generators, directly lowering marginal emissions in those hours \cite{chalamala_data_2025}. However, standard price-based DR mechanisms commonly used for residential and industrial loads are not well aligned with hyperscale AI loads. As noted in \cite{chen_electricity_2025}, sudden rescheduling of deferred AI tasks can create rebound peaks that increase fossil-fuel use and strain the grid, limiting the sustainability benefits DR is intended to provide. In many markets, the financial incentives associated with DR are modest relative to the operational and compliance risks borne by hyperscale operators, further limiting participation. Recognizing this, utilities are beginning to explore incentive- or contract-based DR structures that offer more predictable load relief. For example, Google has made agreements with utilities in Indiana and Tennessee, which explicitly require reducing AI workloads during grid emergencies, thereby helping the system avoid high-emission generation and supporting overall decarbonization goals \cite{michael_terrell_how_2025}. Yet participation remains modest. Operational complexity, reliability concerns, and insufficient financial incentives continue to prevent many AI data centers from engaging in DR at scale, which is an outcome highlighted repeatedly in industry assessments \cite{chalamala_data_2025}. Without addressing these structural barriers, the potential of AI data centers to act as flexible, low-carbon grid resources will remain underutilized. 

\subsection{AI Data Centers as Anchor Lods for Renewable and Transmission Investment }

Beyond their direct, real-time operational interactions with the grid, AI data centers also function as anchor loads that are influencing long-term infrastructure planning and investment. Because such facilities typically require continuous power supply in significant scale and enter into multi-decade service agreements, they provide a form of stable and predictable demand. This long-term demand certainty improves the bankability of renewable energy projects by reducing revenue risk and enabling developers to secure financing at lower cost of capital. In several regions, utilities and independent power producers have noted that large data center commitments can accelerate the construction of renewable plants by serving as dependable off-takers for a substantial portion of project output. This effect has been illustrated by several recent developments. Google, for example, has entered into an advanced clean energy agreement with Kairos Power to support the deployment of small modular reactors (SMRs) that could collectively enable up to 500 MW of clean power during the next decade \cite{michael_terrell_new_2024}, leveraging its future electricity demand to help anchor early-stage nuclear investment. Similarly, Meta’s long-term power purchase agreement with PNM and NextEra supported the commissioning of the 190 MW Sky Ranch Solar Energy Center paired with a 50 MW battery system in New Mexico \cite{pnm_pnm_2024}, a project that local regulators and utilities described as economically viable in part because of the long-term load commitment. In the Middle East, Dubai’s Moro Hub Green Data Center demonstrates another variant of this anchoring dynamic: its large, sustained electricity needs is reported to be powered by a solar facility exceeding 100 MW within the Mohammed bin Rashid Al Maktoum Solar Park \cite{morohubWorldsLargestGreen2025}, reducing project risk while enabling clean, local energy supply at scale. 

The anchoring effect extends beyond generation. Large, concentrated loads can also justify upgrades to high-voltage substations, transformer capacity, and interconnection facilities, particularly in resource-rich areas where renewable potential exceeds the transmission capacity available to deliver it. Multiple US regions now exhibit clear evidence that long-term hyperscale load commitments directly enable major transmission expansions. PJM’s 2023 Regional Transmission Expansion Plan (RTEP) approved \$1.2 billion of 500 kV and 230 kV reinforcements partly driven by the “Data Center Alley”, located in northern Virginia \cite{2023rtepreport}. Entergy Louisiana is constructing a \$1.2 billion, 100-mile high-voltage line to serve Meta’s 2 GW campus \cite{entergy_entergy_2025}, while American Transmission Company (ATC) has proposed a \$1.4 billion, 345 kV line to power the Vantage Data Centers in Port Washington, Wisconsin, a facility proposed with 1.3 GW demand in its first development phase \cite{zotero-item-809}. In each case, the guaranteed multi-GW anchor load improves the economic viability of long-lead transmission investments and shifts project-planning risk away from utilities and regulators. Collectively, these examples illustrate that when coordinated effectively, the co-development of data centers, renewable generation, and transmission infrastructure can enable regions to unlock otherwise stranded renewable resources while improving overall grid reliability. 

However, the degree to which AI data centers can catalyze such investment remains strongly dependent on system context. In regions where transmission corridors are already heavily congested, or where new lines face multi-year permitting, siting, and land-use challenges, even a multi-GW anchor load may not be sufficient to overcome structural bottlenecks that delay grid expansion. Physical constraints such as limited rights-of-way, a saturated interconnection queue, or insufficient availability of suitable land for co-located renewable build-out can further limit the responsiveness of infrastructure development to large load signals. Institutional factors are equally significant. In many jurisdictions, transmission planning follows formalized, multiyear regional processes that do not allow bilateral coordination between individual loads and transmission operators. Cost allocation rules may also prevent targeted investments even when the economic case is strong, while market designs that separate load siting from resource planning limit the influence that hyperscale customers can exert on long-term infrastructure decisions. These constraints suggest that the anchoring effect of AI data centers is inherently conditional: powerful where planning frameworks can integrate large load commitments into forward looking transmission and generation expansion, and muted where regulatory, procedural, or physical barriers impede such integration.

\section{Corporate Sustainability Practices and Pathways}

As AI-driven electricity demand has accelerated, major technology companies have become central actors in shaping corporate clean energy procurement and sustainable data center development. As the operators of majority of the world’s largest hyperscale facilities, these firms have articulated ambitious decarbonization goals. Their strategies include large-scale procurement of renewable electricity, investments in emerging clean energy technologies, deployment of innovative energy management systems, and the redesign of data center infrastructure to reduce operational and embodied footprints. This section reviews these real-world practices and examines their implications for system-level sustainability, while also assessing the limitations of current corporate reporting frameworks. 

\subsection{Corporate commitments and clean energy procurement pathways}

Leading technology companies have positioned sustainability as a central pillar of their corporate strategies, each outlining ambitious climate commitments supported by sizeable clean energy investments. Table \ref{hyperscale_sustainability} provides a comparative overview of key corporate sustainability goals and achievements across major hyperscale operators. Collectively, these companies demonstrate substantial commitment to sustainable operation objectives over the coming decades, including tens of gigawatts in renewable procurement, including large-scale investments in PPAs, sustained efforts to improve data center energy efficiency, alongside the development of many other sustainability mechanisms \cite{meta_2025_2025, google_environment_2025, amazon_2024_2024, microsoft_2025_2025, google_moving_2018}. 

\begin{table*}[t]
\centering{
\begin{threeparttable}
\caption{Sustainability goals and reported practices of major hyperscale operators}
\label{hyperscale_sustainability}
\renewcommand{\arraystretch}{1.3}
\begin{tabular}{p{2.5cm} p{3.2cm} p{3.6cm} p{3.2cm} p{3.0cm}}
\toprule
\textbf{Company} 
& \textbf{Google \cite{google_environment_2025, google_moving_2018, google_growing_nodate}} 
& \textbf{Meta \cite{meta_2025_2025, meta_2024_2024}} 
& \textbf{Microsoft \cite{microsoft_2025_2025, microsoftMeasuringEnergyWater}} 
& \textbf{Amazon \cite{amazon_2024_2024, aws_aws_nodate}} \\
\midrule

\textbf{Sustainability Goal} 
& Net-zero by 2030; 24/7 Carbon-free energy (CFE) 
& Net-zero by 2030 
& Carbon negative by 2030 
& Net-zero by 2040 \\

\textbf{Data center PUE*} 
& 1.09 
& 1.08 
& 1.16 
& 1.15 \\

\textbf{PPA} 
& Over 22 GW by 2024 
& Over 15 GW in the past decade 
& 34 GW by 2024 
& 34 GW by January 2025 \\

\textbf{Annual renewable electricity matching} 
& 100\% since 2017 
& 100\% since 2020 
& Targeting 100\% by 2030 
& 100\% since 2023 \\

\bottomrule
\end{tabular}

\begin{tablenotes}
\footnotesize
\item *Reported PUE values reflect fleet-wide annual averages and are based on the latest publicly disclosed reporting periods, which differ across firms.
\end{tablenotes}

\end{threeparttable}
}
\end{table*}

\subsection{From Annual Matching to granular Carbon Accounting: Practices, Limitations, and Challenges }

Major AI data center operators have made substantial investments in clean energy and report their sustainability performance under prevailing corporate carbon accounting practices. In contemporary corporate decarbonization strategies, virtual PPAs and annual clean energy matching operate as a coupled procurement–accounting system \cite{kansal_introduction_2018}: virtual PPAs supply renewable electricity attributes, while annual matching provides the accounting framework that allocates these attributes to a firm’s electricity consumption on a yearly basis. 

Although the technical mechanisms of virtual PPAs have been introduced earlier in the section ‘Onsite and Procured Clean Energy Resources’, it is important to highlight their historical contribution within this combined system. Over the past decade, virtual PPAs have played a pivotal role in scaling global renewable deployment by providing long-term revenue certainty, reducing project financing risks, and mobilizing gigawatt-scale investment. As shown in Table 1, major hyperscale operators have each contracted tens of gigawatts of renewable capacity annually, an outcome that would have been unlikely without the architectures enabled by virtual PPAs. This procurement pathway has therefore been instrumental in accelerating clean energy deployment, even when the generation is geographically separate from the data centers it nominally supports. 

At the accounting level, annual matching remains the predominant framework used to translate procured renewable attributes into corporate sustainability claims \cite{meta_2025_2025, google_environment_2025, amazon_2024_2024, microsoft_2025_2025}. Under this framework, a company’s total annual renewable procurement is compared with its total annual electricity use, regardless of when or where electricity is actually consumed. Despite its coarse temporal and spatial resolution, annual matching has played an important role in corporate decarbonization pathways. It offers a straightforward, auditable structure that enables multinational firms to report renewable electricity usage consistently across diverse regulatory and market environments. In many regions lacking hourly marginal emissions data or granular grid transparency, annual matching remains the only feasible method for converting clean energy procurement into reportable sustainability outcomes. It also functions as an essential entry point for the sustainable operation of corporates, lowering the threshold for organizations to transition from fossil-based electricity to clean procurement, serving as a precursor to more granular carbon accounting mechanisms, and helping firms establish the foundational elements of internal carbon governance. 

Despite the substantial progress enabled by this coupled mechanism, it also carries important limitations. While useful from a bookkeeping standpoint, annual matching does not capture the hourly or locational carbon intensity of the electricity actually serving data centers. As a result, a company may report “100\% renewable energy” even during hours when its facilities are physically supplied by fossil-fuel marginal generators. A second limitation arises from the geographic decoupling inherent in virtual PPAs, which allow companies to procure clean energy attributes or emissions reductions from regions entirely disconnected from the grids on which their data centers operate. Although such mechanisms continue to support global renewable investment, they do not necessarily reduce fossil-fuel reliance in the specific grids facing the most rapid load growth from data centers. Consequently, firms may achieve paper-based net-zero targets while their local operations continue to drive non-zero emissions, increased curtailment risks, or greater dependence on fossil-based peaking units. This divergence is reflected in reported data: Meta’s 2025 Sustainability Report shows that locational-based operational emissions, which reflect the footprint of the actual grid supplying the data center, can be nearly twice as high as emissions calculated using global energy procurement totals \cite{meta_2025_2025}. 

In recognition of the limitations of annual and geographically aggregated accounting approaches, several leading hyperscale operators have begun moving toward more granular matching mechanisms that better align clean electricity supply with real-time data center demand. Among these emerging approaches, Google’s 24/7 CFE goal \cite{google_moving_2018} represents one of the most advanced formulations, targeting hourly, grid-level matching between electricity consumption and carbon-free generation. Google reported a 66\% hourly CFE score in 2024 \cite{google_environment_2025}. Early-stage industry discussions also have reflected growing interest in moving toward finer granularities in carbon accounting approaches, even if formal commitments remain limited. 

However, implementing 24/7 CFE at scale presents substantial practical and structural challenges. 
\begin{enumerate}
    \item Data availability. Hourly matching requires high-resolution, real-time grid carbon-intensity and emissions data and transparency into marginal generation resources, which is not yet available or standardized across many electricity markets.  
    \item Tracking system. Interoperable tracking systems for hourly energy attributes remain in early development, and the absence of standardized methodologies complicates verification. 
    \item Resource portfolio requirements. Achieving hourly carbon-free supply often requires procuring from a more diverse portfolio of clean energy resources, increasing procurement complexity and transaction costs. 
    \item Market and regulatory readiness. The market design and regulatory frameworks in many regions do not yet support the granular attribution, certification, or trading of time-stamped clean energy attributes. 
    \item Physical grid constraints. Constraints such as transmission congestion and regional clean energy scarcity during certain hours limit the practical feasibility of achieving high hourly CFE scores in some locations. 
\end{enumerate}

Together, these factors suggest that while granular matching mechanisms such as 24/7 CFE represent a promising evolution beyond annual accounting, their widespread adoption will depend on parallel advances in data availability, standards development, market design, and grid infrastructure. 

\section{Conclusion}

AI data centers have emerged as one of the most influential forces shaping the sustainability of modern electric power systems. Their unprecedented scale, rapid expansion, and spatial concentration introduce substantial sustainability considerations because they interact with regional generation mixes, transmission constraints, and planning processes in ways that can influence system conditions. In regions where clean energy deployment and transmission expansion cannot keep pace with aggregate load growth from multiple sources, sustained increases in data center demand may contribute to conditions under which marginal emissions remain elevated, or fossil units remain viable for longer. These outcomes reflect broader system dynamics rather than the characteristics of data centers alone. The environmental implications of AI-driven electricity use are therefore best understood as context-dependent, shaped by regional infrastructure, institutional arrangements, and planning horizons. 

A key insight arises from the flexibility potential of AI workloads. While flexibility can provide system benefits, its realizable magnitude is highly context-dependent and should not be assumed universal. AI data centers can also act as anchor loads for new renewable and transmission investment, yet this effect is highly dependent on permitting environments, interconnection capacity, and institutional planning structures, and therefore cannot be generalized across markets. Corporate procurement practices offer another important lens. Virtual PPAs and annual matching have played a foundational role in scaling renewable deployment, but their environmental representativeness is limited by their lack of temporal and locational specificity, and more granular approaches face practical and regulatory hurdles. 

Taken together, these observations underscore that the sustainability impacts of AI data centers are strongly scenario and context dependent, and effective solutions will need to be tailored to regional grid conditions, planning institutions, and market designs rather than assuming a one size fits all approach. Looking ahead, both research and industry practice must shift toward co-designing AI data centers and power systems. With thoughtful coordination, AI data centers can evolve from emerging sustainability risks into active contributors to a low-carbon, resilient electric grid.

\bibliographystyle{IEEEtran}
\bibliography{references}

\end{document}